\documentclass{iaus}

\title[Basic Space Science TRIPOD]
{The United Nations Basic Space Science Initiative: The TRIPOD concept}

\author[Kitamura et al.]
{M. Kitamura$^1$, D. Wentzel$^2$, A.A. Henden$^3$, J. Bennett$^4$, H.M.K. Al-Naimiy$^5$, A.M. Mathai$^6$, N. Gopalswamy$^7$,
J. Davila$^8$, B. Thompson$^9$, D.F. Webb$^{10}$, \break \and H.J. Haubold$^{11}$
 \thanks{Presented the invited paper on behalf of the United Nations Office for Outer Space Affairs}}

\affiliation{$^1$National Astronomical Observatory, Mitaka, Tokyo 181-8588, Japan \break \\[\affilskip]
$^2$University of Maryland, College Park, MD 20742-2421, USA \break email: d.wentzel@worldnet.att.net\\[\affilskip]
$^3$American Association of Variable Star Observers, 25 Birch Street, Cambridge, MA 02138, USA \break email: arne@aavso.org\\[\affilskip]
$^4$3015 10th St., Boulder, CO 80304, USA \break email: jeffrey.bennett@comcast.net\\[\affilskip]
$^5$College of Arts and Sciences, Sharjah University P.O. Box, 27272 Sharjah, United Arab Emirates \break email: alnaimiy2@yahoo.com\\[\affilskip]
$^6$Centre for Mathematical Sciences, Pala Campus, Arunapuram P.O. Box, Pala-686574, Kerala, India \break email: mathai@math.mcgill.ca\\[\affilskip]
$^7$NASA Goddard Space Flight Center, Greenbelt, MD 20771, USA \break email: gopals@ssedmail.gsfc.nasa.gov\\[\affilskip]
$^8$NASA Goddard Space Flight Center, Greenbelt, MD 20771, USA \break email: joseph.m.davila@nasa.gov\\[\affilskip]
$^9$NASA Goddard Space Flight Center, Greenbelt, MD 20771, USA \break email: barbara.j.thompson@nasa.gov\\[\affilskip]
$^{10}$AFRL/VSBXS and ISR, Boston College, 29 Randolph Road, Hanscom AFB, MA 01731-3010, USA \break email: David.Webb.ctr@hanscom.af.mil\\[\affilskip]
$^{11}$United Nations Office for Outer Space Affairs, Vienna International Centre, A-1400 Vienna, Austria \break email: hans.haubold@unvienna.org\\}

\pubyear{2007}
\volume{xxx}  
\pagerange{1--8}
\date{?? and in revised form ??}
\jname{Proceedings Title IAU Symposium}
\editors{A.C. Editor, B.D. Editor \& C.E. Editor, eds.}
\begin{document}

\maketitle

\begin{abstract}
Since 1990, the United Nations is annually holding a workshop on basic space science for the benefit of the worldwide development of astronomy. Additional to the scientific benefits of the workshops and the strengthening of international cooperation, the workshops lead to the establishment of astronomical telescope facilities through the Official Development Assistance (ODA) of Japan. Teaching material, hands-on astrophysics material, and variable star observing programmes had been developed for the operation of such astronomical telescope facilities in an university environment. This approach to astronomical telescope facility, observing programme, and teaching astronomy has become known as the basic space science TRIPOD concept. Currently, a similar TRIPOD concept is being developed for the International Heliophysical Year 2007, consisting of an instrument array, data taking and analysis, and teaching space science.
\end{abstract}

\firstsection
\section{Introduction}

Research and education in astronomy and astrophysics are an international enterprise and the astronomical community has long shown leadership in creating international collaborations and cooperation: Because (i) astronomy has deep roots in virtually every human culture, (ii) it helps to understand humanity's place in the vast scale of the universe, and (iii) it teaches humanity about its origins and evolution.  Humanity's activity in the quest for the exploration of the universe is reflected in the history of scientific institutions, enterprises, and sensibilities. The institutions that sustain science; the moral, religious, cultural, and philosophical sensibilities of scientists themselves; and the goal of the scientific enterprise in different regions on Earth are subject of intense study (Pyenson and Sheets-Pyenson 1999).     

The Decadal Reports for the last decade of the 20th century (Bahcall, 1991) and the first decade of the 21st century (McKee and Taylor, 2001) have been prepared primarily for the North American astronomical community, however, it may have gone unnoticed that these reports had also an impact on a broader international scale, as the reports can be used, to some extend, as a guide to introduce basic space science, including astronomy and astrophysics, in nations where this field of science is still in its infancy. Attention is drawn to the world-wide-web sites at http://www.seas.columbia.edu/$\sim$ah297/\\un-esa/ and http://www.unoosa.org/oosa/en/SAP/bss/index.html, where the TRIPOD concept is publicized on how developing nations are making efforts to introduce basic space science into research and education curricula at the university level. The concept, focusing on astronomical telescope facilities in developing nations, was born in 1990 as a collaborative effort of developing nations, the European Space Agency (ESA), the United Nations (UN), and the Government of Japan. Through annual workshops and subsequent follow-up projects, particularly the establishment of astronomical telescope facilities, this concept is gradually bearing results in the regions of Asia and the Pacific, Latin America and the Caribbean, Africa, and Western Asia (Wamsteker et al. 2004). 

\section{United Nations Office for Outer Space Affairs (UNOOSA)}

In 1959, the United Nations recognized a new potential for international cooperation and formed a permanent body by establishing the Committee on the Peaceful Uses of Outer Space (COPUOS). In 1970, COPUOS formalized the UN Programme on Space Applications to strengthen cooperation in space science and technology between developing and industrialized nations.  

The overall purpose of the programme "Peaceful Use of Outer Space" is the promotion of international cooperation in the peaceful uses of outer space for economic, social and scientific development, in particular for the benefit of developing nations. The programme aims at strengthening the international legal regime governing outer space activities to improve conditions for expanding international cooperation in the peaceful uses of outer space. The implementation of the programme will strengthen efforts at the national, regional and global levels, including among entities of the United Nations system, to increase the benefits of the use of space science and technology for sustainable development.

Within the secretariat of the United Nations, the programme is implemented by the Office for Outer Space Affairs. At the inter-governmental level, the programme is implemented by the Committee on the Peaceful Uses of Outer Space, which addresses scientific and technical as well as legal and policy issues related to the peaceful uses of outer space. The Committee was established by the General Assembly in 1959 and has two subsidiary bodies, the Legal Subcommittee and the Scientific and Technical Subcommittee. The direction of the programme is provided in the annual resolutions of the General Assembly and decisions of the Committee and its two Subcommittees.

As part of its programme of work, the Office provides secretariat services to the Committee and its subsidiary bodies and implements the United Nations Programme on Space Applications. The activities of the Programme on Space Applications are primarily designed to build the capacity of developing nations to use space applications to support their economic and social development.

In its resolution 54/68 of 6 December 1999, the United Nations General Assembly endorsed the resolution entitled "The Space Millennium: Vienna Declaration on Space and Human Development", which had been adopted by the Third United Nations Conference on the Exploration and Peaceful Uses of Outer Space (UNISPACE III), held in July 1999. Since then, the focus of the work undertaken by the Office under this programme has been to assist the Committee in the implementation of the recommendations of UNISPACE III.

In October 2004, the United Nations General Assembly reviewed the progress made in the implementation of the recommendations of UNISPACE III and, in its resolution 59/2, endorsed the Committee's Plan of Action for their further implementation. The Plan of Action, contained in the report of the Committee to the Assembly for its review (A/59/174), constitutes a long-term strategy for enhancing mechanisms to develop or strengthen the use of space science and technology to support the global agendas for sustainable development. The report also provides a road map to make space tools more widely available by moving from the demonstration of the usefulness of space technology to an operational use of space-based services. 

In its report, the Committee noted that in implementing the Plan of Action, the Committee could provide a bridge between users and potential providers of space-based applications and services by identifying needs of Member States and coordinating international cooperation to facilitate access to the scientific and technical systems that might meet them. To maximize the effectiveness of its resources, the Committee adopted a flexible mechanism, action teams, that takes advantage of partnerships among its secretariat, Governments, and intergovernmental and international non-governmental organizations to further implement the recommendations of UNISPACE III.

At its forty-ninth session, held in June 2006, the Committee had before it for its consideration the proposed Strategic Framework for the Office for Outer Space Affairs for the period 2008 - 2009, as contained in document (A/61/6 (Prog.5)). The Committee agreed on the proposed strategic framework.

The expected accomplishments and the strategy reflected in the strategic framework proposed by the Office for Outer Space Affairs for the period 2008-2009 (A/61/6) are aimed at achieving increased international cooperation among Member States and international entities in the conduct of space activities for peaceful purposes and the use of space science and technology and their applications towards achieving internationally agreed sustainable development goals.

In brief, the three expected accomplishments of the Office are: (a) greater understanding, acceptance, and implementation by the international community of the legal regime established by the United Nations to govern outer space activities; (b) strengthened capacities of countries in using space science and technology and their applications in areas related, in particular, to sustainable development, and mechanisms to coordinate their space-related policy matters and space activities; and (c) increased coherence and synergy in the space-related work of entities of the United Nations system and international space-related entities in using space science and technology and their applications as tools to advance human development and increase overall capacity development. The establishment and operation of Regional Centres for Space Science and Technology, affiliated to the United Nations\\
(http://www.unoosa.org/oosa/en/SAP/centres/index.html),\\
as well as workshops on basic space science\\
(http://www.unoosa.org/oosa/en/SAP/bss/index.html)\\
and the International Heliophysical Year 2007\\
(http://www.unoosa.org/oosa/en/SAP/bss/ihy2007/index.html)\\
are part of the accomplishments of the Office.

\section{Official Development Assistance (ODA) of the Government of Japan}

In conjunction to the workshops, to support research and education in astronomy, the Government of Japan has donated high-grade equipment to a number of developing nations (Singapore 1987, Indonesia 1988, Thailand 1989, Sri Lanka 1995, Paraguay 1999, the Philippines 2000, Chile 2001) within the scheme of ODA of the Government of Japan (Kitamura 1999). Here, reference is made to 45cm high-grade astronomical telescopes furnished with photoelectric photometer, computer equipment, and spectrograph (or CCD). After the installation of the telescope facility by the host country and Japan, in order to operate such high-grade telescopes, young observatory staff members from the host country have been invited by the Bisei Astronomical Observatory for education and training, sponsored by the Japan International Cooperation Agency [JICA] (Kitamura 1999, Kogure 1999, Kitamura 2004, UN document A/AC.105/829). Similar telescope facilities, provided by the Government, were inaugurated in Honduras (1997) and Jordan (1999).

The research and education programmes at the newly established telescope facilities focus on time-varying phenomena of celestial objects. The 45cm class reflecting telescope with photoelectric photometer attached is able to detect celestial objects up to the 12th magnitude and with a CCD attached up to the 15th magnitude, respectively. Such results have been demonstrated for the light variation of the eclipsing close binary star V505 Sgr, the X-ray binary Cyg X-1, the eclipsing part of the long-period binary eps Aur, the asteroid No.45 Eugenia, and the eclipsing variable RT CMa (Kitamura 1999).

Also in 1990, the Government of Japan through ODA, facilitated the provision of planetariums to developing nations (Kitamura 2004; Smith and Haubold 1992).

\section{Observing with the Telescopes: Research}

In the course of preparing the establishment of the above astronomical telescope facilities, the workshops made intense efforts to identify available material to be used in research and education by utilizing such facilities. It was discovered that variable star observing by photoelectric or CCD photometry can be a prelude to even more advanced astronomical activity. Variable stars are those whose brightness, colour, or some other property varies with time. If measured sufficiently carefully, almost every star turns out to be variable. The variation may be due to geometry, such as the eclipse of one star by a companion star, or the rotation of a spotted star, or it may be due to physical processes such as pulsation, eruption, or explosion. Variable stars provide astronomers with essential information about the internal structure and evolution of the stars. The most preeminent institution in this specific field of astronomy is the American Association of Variable Star Observers. The AAVSO co-ordinates variable star observations made by amateur and professional astronomers, compiles, processes, and publishes them, and in turn, makes them available to researchers and educators.   

To facilitate the operation of variable star observing programmes and to prepare a common ground for such programmes, the AAVSO developed a rather unique package titled ``Hands-On Astrophysics'' which includes 45 star charts, 31 35mm slides of five constellations, 14 prints of the Cygnus star field at seven different times, 600,000 measurements of several dozen stars, user-friendly computer programmes to analyze them, and to enter new observations into the database, an instructional video in three segments, and a very comprehensive manual for teachers and students (http://www.aavso.org/). Assuming that the telescope is properly operational, variable stars can be observed, measurements can be analyzed and sent electronically to the AAVSO. 

The flexibility of the ``Hands-On Astrophysics'' material allows an immediate link to the teaching of astronomy or astrophysics at the university level by using the astronomy, mathematics, and computer elements of this package. It can be used as a basis to involve both the professor and the student to do real science with real observational data. After a careful exploration of ``Hands-On Astrophysics'' and thanks to the generous cooperation of the AAVSO, it was adopted by the above astronomical telescope facilities for their observing programmes (Mattei and Percy 1999, Percy 1991, Wamsteker et al. 2004).

The AAVSO is currently undertaking a massive effort to translate its basic Visual Observing Manual into many languages such as Spanish and Russian, to make this basic material available in the native language of any developing nation. The AAVSO is actively pursuing translations in Arabic and Chinese so as to have versions available in all the official United Nations languages.

\section{Teaching Astrophysics: Education}

Various strategies for introducing the spirit of scientific inquiry to universities, including those in developing nations, have been developed  and analyzed (Wentzel 1999a). The workshops on basic space science were created to foster scientific inquiry. Organized and hosted by Governments and scientific communities, they serve the need to introduce or further develop basic space science at the university level, as well as to establish adequate facilities for pursuing a scientific field in practical terms. Such astronomical facilities are operated for the benefit of the university or research establishment, and will also make the results from these facilities available for public educational efforts. Additional to the hosting of the workshops, the Governments agreed to operate such a telescope facility in a sustained manner with the call on the international community for support and cooperation in devising respective research and educational programmes.  

Organizers of the workshops have acknowledged in the past the desire of the local scientific communities to use educational material adopted and available at the local level (prepared in the local language). However, the workshops have also recommended to explore the possibility to develop educational material (additional to the above mentioned ``Hands-On Astrophysics'' package) which might be used by as many as possible university staff in different nations while preserving the specific cultural environment in which astronomy is being taught and the telescope is being used. A first promising step in this direction was made with the project ``Astrophysics for University Physics Courses'' (Wentzel 1999b, Wamsteker et al. 2004). This project has been highlighted at the IAU/COSPAR/UN Special Workshop on Education in Astronomy and Basic Space Science, held during the UNISPACE III Conference at the United Nations Office Vienna in 1999 (Isobe 1999). Additionally, a number of text books and CD-ROMs have been reviewed over the years which, in the view of astronomers from developing nations, are particularly useful in the research and teaching process: Bennett et al. 2007, for teaching purposes and Bennett, 2001, Lang, 1999, 2004 reference books in the research process.

As part of the 15th anniversary celebrations of the Hubble Space Telescope, the European Space Agency has produced an exclusive, 83-minute DVD film, entitled "Hubble -- 15 Years of Discovery". The documentary also mentions the role of the Hubble Space Telescope project in facilitating some of the activities of the United nations Office for Outer Space Affairs, particularly processing of Hubble imagery as part of the education and research activities of the UN-affiliated Regional Centres for Space Science and Technology and the workshops on basic space science. The Hubble DVD was distributed world-wide, through the Office, as a unique educational tool for astronomy and astrophysics.

\section{International Heliophysical Year 2007: A World-Wide Outreach Programme}

In 1957 a programme of international research, inspired by the International Polar Years of 1882-83 and 1932-33, was organized as the International Geophysical Year (IGY) to study global phenomena of the Earth and geospace. The IGY involved about 66,000 scientists from 60 nations, working at thousands of stations, from pole to pole to obtain simultaneous, global observations on Earth and in space. The fiftieth anniversary of IGY will occur in 2007. It was proposed to organize an international programme of scientific collaboration for this time period called the International Heliophysical Years (IHY) in 2007 (http://ihy2007.org/). Like IGY, and the two previous International Polar Years, the scientific objective of IHY is to study phenomena on the largest possible scale with simultaneous observations from a broad array of instruments. Unlike previous international years, today observations are routinely received from a vast armada of sophisticated instruments in space that continuously monitor solar activity, the interplanetary medium, and the Earth. These spacecraft together with ground level observations and atmospheric probes provide an extraordinary view of the Sun, the heliosphere, and their influence on the near-Earth environment. The IHY is a unique opportunity to study the coupled Sun-Earth system. Future basic space science workshops will focus on the preparation of IHY 2007 world-wide, particularly taking into account interests and contributions from developing nations.

Currently, in accordance with the United Nations General Assembly resolution 60/99, the Scientific and Technical Subcommittee of the UNCOPUOS is considering an agenda item on the IHY 2007 under the three-year work plan adopted at the forty-second session of the Subcommittee\\
(http://www.unoosa.org/oosa/en/SAP/bss/ihy2007/index.html).\par

A major thrust of the IHY 2007 is to deploy arrays of small, inexpensive instruments such as magnetometers, radio antennas, GPS receivers, all-sky cameras, etc. around the world to provide global measurements of ionospheric, magnetospheric, and heliospheric phenomena. This programme is implemented  by collaboration between the IHY 2007 Secretariat and the United Nations Office for Outer Space Affairs. The small instrument programme consists of a partnership between instrument providers and instrument host nations. The lead scientist or engineer provides the instrumentation (or fabrication plans for instruments) in the array; the host nation provides manpower, facilities, and operational support to obtain data with the instrument, typically at a local university. In preparation of IHY 2007, this programme has been active in deploying instrumentation, developing plans for new instrumentation, and identifying educational opportunities for the host nation in association with this programme\\ (http://ihy2007.org/observatory/observatory.shtml;\\
UN document A/AC.105/856). Currently, a TRIPOD concept is being developed for the International Heliophysical Year 2007, consisting of an instrument array, data taking and analysis, and teaching space science.

\section{Remark}

In 2006, 27 November - 1 December, the Indian Institute of Astrophysics will host the second UN/NASA Workshop on the International Heliophysical Year and Basic Space Science in Bangalore, India (http://www.iiap.res.in/ihy/). In 2007, 11-15 June, the National Astronomical Observatory of Japan, Tokyo, will host a workshop on basic space science and the International Heliophysical Year 2007, co-organized by the United Nations, European Space Agency, and the National Aeronautics and Space Administration of the United States of America, and will use this opportunity to commemorate the cooperation between the Government of Japan and the United Nations, as highlighted in this article, since 1990.\par
\bigskip
\noindent 
{\bf References}\par
\bigskip
\noindent
Bahcall, J., The Decade of Discovery in Astronomy and Astrophysics, National Academy Press, Washington D.C., 1991; and Astronomy and Astrophysics: Panel Reports, National Academy Press, Washington D.C., 1991.\par
\smallskip
\noindent
Bennett, J., On the Cosmic Horizon: Ten Great Mysteries for the Third Millennium Astronomy, Addison Wesley Longman 2001.\par
\smallskip
\noindent
Bennett, J., Donahue, M., Schneider, N., and Voit, M., The Cosmic Perspective, Addison Wesley Longman Inc., Menlo Park, California, Fourth Edition, 2007; CD-ROMs and a www site, offering a wealth of additional material for professors and students, specifically developed for teaching astronomy with this book and upgraded on a regular basis are also available: http://www.masteringastronomy.com/.\par
\smallskip
\noindent
Haubold, H.J., ``UN/ESA Workshops on Basic Space Science: an initiative in the world-wide development of astronomy'',  Journal of Astronomical History and Heritage 1(2):105-121, 1998; Space Policy 19:215-219, 2003.\par
\smallskip
\noindent 
Isobe, S., Teaching of Astronomy in Asian-Pacific Region, Bulletin No. 15, 1999.\par
\smallskip
\noindent
McKee, C.F. and Taylor, Jr., J.H., Astronomy and Astrophysics in the New Millennium, National Academy Press, Washington D.C., 2001; and Astronomy and Astrophysics in the New Millennium: Panel Reports, National Academy Press, Washington D.C., 2001; see also G. Brumfield, Wishing for the stars, Nature 443(2006)386-389.\par
\smallskip
\noindent 
Kitamura, M., ``Provision of astronomical instruments to developing countries by Japanese ODA with emphasis on research observations by donated 45cm reflectors in Asia'', in Conference on Space Sciences and Technology Applications for National Development: Proceedings, held at Colombo, Sri Lanka, 21-22 January 1999, Ministry of Science and Technology of Sri Lanka, pp. 147-152.\par
\smallskip
\noindent
Kitamura, M., "Aiding astronomy in developing nations: Japanese ODA", Space Policy, 20:131-135, 2004.\par 
\smallskip
\noindent 
Kogure, T., ``Stellar activity and needs for multi-site observations'', in Conference on Space Sciences and Technology Applications for National Development: Proceedings, held at Colombo, Sri Lanka, 21-22 January 1999, Ministry of Science and Technology of Sri Lanka, pp. 124-131.\par
\smallskip
\noindent 
Lang, K.R., Astrophysical Formulae, Volume I: Radiation, Gas Processes and High Energy Astrophysics, Volume II: Space, Time, Matter and Cosmology, Springer-Verlag, Berlin 1999.\par
\smallskip
\noindent
Lang, K.R., An education curriculum for space science in developing countries, Space Policy 20:297-302, 2004.\par
\smallskip
\noindent 
Mattei, J. and Percy, J. R. (Eds.), Hands-On Astrophysics, American Association of Variable Star Observers, Cambridge, MA 02138, 1998;\\ http://www.aavso.org/.\par
\smallskip
\noindent 
Percy, J.R. (Ed.), Astronomy Education: Current Developments, Future Cooperation: Proceedings of an ASP Symposium, Astronomical Society of the Pacific Conference Series Vol. 89, 1991.\par
\smallskip
\noindent 
Pyenson, L. and Sheets-Pyenson, S., Servants of Nature: A History of Scientific Institutions, Enterprises, and Sensibilities, W.W. Norton \& Company, New York, 1999.\par
\smallskip
\noindent
Smith, D.W. and Haubold, H.J. (Eds.): Planetarium: A Challenge for Educators, United Nations, New York 1992; available in Japanese, English, Spanish, and Slovak languages.\par
\smallskip
\noindent
UN document A/AC.105/829: Report on the Twelfth United Nations/European Space Agency Workshop on Basic Space Science, Beijing, China, 24-28 May 2004, United Nations, Vienna 2004.\par
\smallskip
\noindent
UN document A/AC.105/856: Report on the United Nations/European Space Agency/National Aeronautics and Space Administration of the United States of America Workshop on the International Heliophysical Year 2007, Abu Dhabi and Al-Ain, United Arab Emirates, 20-23 November 2005, United Nations, Vienna 2005.\par
\smallskip
\noindent
Wamsteker, W., Albrecht, R., and Haubold, H.J. (Eds.): Developing Basic Space Science World-Wide: A Decade of UN/ESA Workshops, Kluwer Academic Publishers, Dordrecht 2004.\par
\smallskip
\noindent 
Wentzel, D.G., ``National strategies for science development'', Teaching of Astronomy in Asian-Pacific Region, Bulletin No. 15, 1999a, pp. 4-10.\par
\smallskip
\noindent 
Wentzel, D.G., Astrofisica para Cursos Universitarios de Fisica, La Paz, Bolivia, 1999b, English language version available from the United Nations in print and electronically at\\ http://www.seas.columbia.edu/$\sim$ah297/un-esa/astrophysics; printed version also contained in Wamsteker at al. 2004.
\end{document}